%% file: report.tex
\documentclass[letterpaper,10 pt,conference]{ieeeconf}

\IEEEoverridecommandlockouts
\usepackage{graphics,epsfig,mathptmx,times,amsmath,amssymb,graphicx,booktabs}
\usepackage[ruled,vlined]{algorithm2e}
\usepackage[english]{babel}
\usepackage{blindtext}

\input defs.tex

\title{\LARGE \bf
Player Behavior and Optimal Team Composition in\\
Online Multiplayer Games
}

\author{Hao Yi Ong$^{1}$, 
        Sunil Deolalikar$^{2}$ and 
        Mark V. Peng$^{3}$%
\thanks{$^{1}$Mechanical Engineering Department, 
        Stanford University}%
\thanks{$^{2}$Aeronautics and Astronautics Department, 
        Stanford University}%
\thanks{$^{3}$Computer Science Department, 
        Stanford University}%
\thanks{Email: {\tt\small $\{$haoyi,sunild93,mvpeng$\}$@stanford.edu}}
}

\begin{document}

\maketitle
\thispagestyle{plain}
\pagestyle{plain}

\begin{abstract} 

We consider clustering player behavior and learning the optimal team composition for multiplayer online games. The goal is to determine a set of descriptive play style groupings and learn a predictor for win/loss outcomes. The predictor takes in as input the play styles of the participants in each team; i.e., the various team compositions in a game. Our framework uses unsupervised learning to find behavior clusters, which are, in turn, used with classification algorithms to learn the outcome predictor. For our numerical experiments, we consider League of Legends, a popular team-based role-playing game developed by Riot Games. We observe the learned clusters to not only corroborate well with game knowledge, but also provide insights surprising to expert players. We also demonstrate that game outcomes can be predicted with fairly high accuracy given team composition-based features.

\end{abstract}

\begin{keywords}

team performance, team composition, player behavior, video games, multiplayer games, game prediction

\end{keywords}

\section{Introduction}

Online virtual worlds are an increasingly significant venue for human interaction. By far the most active virtual worlds belong to a genre of video games called massively multiplayer online role-playing games (MMORPGs), where players interact with each other in a virtual world \cite{TSF:13}. In an MMORPG, players assume the role of in-game characters and take control over most of their characters' actions, often working in teams to accomplish a common objective, such as defeating opposing teams. Due to the shared, persistent nature of these virtual worlds, user behaviors and experiences are shaped by various social factors. 

Besides profit-making, an understanding of these social dynamics would provide insight to human interactions in the real world and the potential of virtual worlds for education, training, and scientific research \cite{Bai:12,Dic:05}. Numerous prior studies in social sciences and management have investigated how team compositions can affect team performance \cite{Spo:11,HA:08}. However, little is understood about player behavior and team performance and factors contributing to it in competitive MMORPGs. To address this, we develop a machine learning framework that uses game histories to learn player behavior clusters and predict the outcome of games given prior knowledge about the game and its players. 

The contributions of this paper are twofold. First, we present several approaches that group player behaviors in online games. Second, we develop predictors that determine how likely it is that a team of players can emerge victorious given the said team's composition of players, all of whom may have different play styles. Specifically, we consider k-means and DP-means---an expectation maximization algorithm \cite{KJ:12}---for clustering play styles and logistic regression (LR), Gaussian discriminant analysis (GDA), and support vector machines (SVMs) for determining win/loss outcomes. The rest of the paper is structured as follows. Section~II describes the target game of our numerical experiments and our data collection method. Sections~III and IV demonstrate several methods and their effectiveness for learning play style clusters and outcome predictors. Some concluding remarks are drawn and future works mentioned in Section~V.

\section{Target Game Description}

We begin with a description of the MMORPG used for our numerical experiments and the data acquisition method.

\subsection{League of Legends}

For this project we consider a popular MMORPG---the League of Legends (LoL). LoL is a multiplayer online battle arena video game developed and published by Riot Games with 27 million daily players \cite{Tas:14}. Furthermore, LoL is a representative MMORPG of its genre, with many similar counterparts such as World of Warcraft's Dota 2 \cite{Suz:09}---giving us a measure of generalizability to other games in its genre. In this MMORPG, a standard game consists of two opposing teams of five players. Each player assumes the role of one of over 120 different characters battling each other to destroy the opposing team's ``towers''---structures that fall after suffering enough attacks from characters. A game is won when all of either team's towers are destroyed. 

\subsection{Data set acquisition}

The developer of LoL has made the game's player statistics and match histories freely available through a web-based application programming interface (API) \cite{Rio:14}. We randomly gathered over 100,000 instances of player statistics and over 10,000 instances of match histories from the 2013-2014 season. We then parsed and cleaned the raw game data to construct our training and testing sets, depending on the features we chose. Player statistics include performance indicators such as average damage dealt and number of wins. Match histories contain information such as participant ID numbers and character choices.

\section{Behavioral Clustering}

The target game's developers have grouped the 120 different in-game characters into six classes, such as assassin or support, which indicates the character's gameplay style. While these classes reflect the developers' design intent for the characters, they do not necessarily reveal the behavior of actual players in games. Using statistics from various players, we present our feature selection method and the gameplay styles learned by applying various clustering algorithms to our data set. We validate our results and the insights derived from it with expert analysis from ranked players.

\subsection{Feature selection}

For our clustering algorithms, the features were 21 normalized player statistics, such as average damage dealt and money earned. The statistics were normalized over their range of values, preventing clusters from being formed due to order of magnitude differences between statistics. For instance, damage dealt values are often 7 orders of magnitude greater than kill streaks, which means small variations in damage dealt are erroneously considered as much more important than kill streaks if taken directly as feature values. 

\subsection{Clustering models}

\subsubsection{k-means}

Given a set of observations, k-means clustering aims to partition them into k sets $\mathcal{S} = \left\{ S_{1},\ldots,S_{k} \right\}$ so as to minimize the within-cluster sum of squares; i.e., find the minimizer $\mathcal{S}^{\star}$ of the distortion function:
\begin{align}
  \sum_{i=1}^{k}\sum_{x \in S_{i}}\left\| x - \mu_{i} \right\|_{2}^{2},
  \label{eqn:k-means}
\end{align}
where $x$ is an observation and $\mu_{i}$ is the $i^{th}$ cluster centroid.

In general, this problem is computationally difficult (NP-hard). For our clustering, we employ Lloyd's algorithm, which is a heuristic that consists of randomly choosing observations as cluster centroids and iteratively assigning observations to their closest centroids and updating the centroids with the mean of their respective clusters \cite{Ng:14}. 

To select the number of clusters $k$, we run 10-fold cross validation over $k$ to find a local optimizer. The scoring function for the cross validation is simply the average distortion given by \eqref{eqn:k-means} over the held-out sets. 

\subsubsection{DP-means}

DP-means is a nonparametric expectation-maximization (EM) algorithm derived using a Dirichlet process (DP) mixture of Gaussians model, which  In other words, the user does not choose the number of clusters beforehand. The technique being the topic of a series of papers, we will only provide a brief description of the algorithm. The reader is referred to \cite{TKJ:13,KJ:12} for a thorough review of DP-means. 

Recall that the standard mixture of Gaussians assumes that one chooses a cluster with probability $\pi_{c}$ and then generates an observation from the $k$ Gaussians corresponding to that chosen cluster. In contrast, the DP mixture of Gaussians is a Bayesian extension to this model that arises by first placing a Dirichlet prior $\text{Dir}(k,\pi_{0})$ on the $k$ mixing Gaussian coefficients (i.e., the probability of choosing a cluster) for some initial set of coefficients $\pi_{0}$ (e.g., uniform prior). As observations are made, the prior is updated and the mixture coefficients change to reflect these new knowledge.

The derivation of DP-means is inspired by the connection between k-means EM with a finite mixture of Gaussians model. Namely, the k-means algorithm may be viewed as a limit of the EM algorithm if all of the covariance matrices corresponding to the clusters in a Gaussian mixture model are equal to $\sigma I$. As $\sigma \rightarrow 0$, the negative log-likelihood of the mixture of Gaussians model approaches the k-means clustering objective \eqref{eqn:k-means}. Correspondingly, the EM steps approach the k-means steps in Lloyd's algorithm. 

In the case of DP-means, \cite{KJ:12} shows how to perform a similar limiting argument. Specifically, suppose that the generative model for the EM algorithm was a DP mixture of Gaussians model with covariances equal to $\sigma I$. Letting $\sigma \rightarrow 0$ for the DP mixture model yields the objective function
\begin{align}
  \sum_{i = 1}^{k}\sum_{x \in S_{i}} \left\| x - \mu_{i} \right\|_{2}^{2} + \left(k - 1\right)\lambda^{2},
  \label{eqn:dp-means}
\end{align}
where $\mathcal{S} = \left\{ S_{1},\ldots,S_{k} \right\}$ is the set of clusters, $x$ is an observation, and $\mu_{i}$ is the $i^{th}$ cluster centroid. Note that, unlike in k-means, $k$ is now a variable to be optimized over.

This leads to an algorithm with clustering assignments similar to the classical k-means algorithm and the same monotonic local convergence guarantees. (See Algorithm~\ref{alg:dp-means}.) The difference is that a new cluster is formed whenever an observation is sufficiently far away from all existing cluster centroids, with some user-defined threshold distance $\lambda$. Intuitively, $\lambda$ is a penalty on the number of clusters, on top of the original k-means distortion function. 

\begin{algorithm}[!htbp]
  \caption{DP-means}

  \DontPrintSemicolon

  \SetKwInOut{Input}{input}
  \SetKwInOut{Output}{output}
  
  \Input{$\mathcal{X}$: input data, $\lambda$: threshold distance}
  \Output{Clustering $S_{1},\ldots,S_{k}$, number of clusters $k$}
  
  \smallskip

  $k \leftarrow 1$ \\
  $S_{1} \leftarrow$ random observation $\left\{x^{\text{rand}}\in\mathcal{X}\right\}$ \\
  $\mu_{1} \leftarrow x^{\text{rand}}$

  \Repeat{
    $S_{1},\ldots,S_{k}$ converge
  }{
    $X^{\text{perm}} \leftarrow$ random ordered permutation of $\mathcal{X}$ \\
    // cluster assignments \\
    \For{
      $x \in X^{\text{perm}}$ in order
    }{
      $c \leftarrow \argmin_{i \in \left\{1,\ldots,k\right\}} \left\| x - \mu_{i} \right\|_{2}^{2}$ \\
      \If{
        $\left\| x - \mu_{c} \right\|_{2}^{2} > \lambda^{2}$
      }{
        $k \leftarrow k + 1$ \\ 
        $\mu_{k} \leftarrow x$
      }
      \Else{
        $\mathcal{S}_{c} \leftarrow \mathcal{S}_{c} \cup \left\{ x \right\}$
      }
    }
    // centroid updates \\
    \For{
      $i = 1,\ldots,k$
    }{
      $\mu_{i} \leftarrow \frac{1}{\left|S_{i}\right|} \sum_{x \in S_{i}} x$
    }
  }

  \label{alg:dp-means}
\end{algorithm}

We ran DP-means with 10-fold cross validation over a range of $\lambda$ values, setting our scoring function as the average of the objective values from \eqref{eqn:dp-means} over the held-out sets.

\subsection{Numerical results}

\begin{figure}[htbp!]
  \centering
  \includegraphics[trim=55pt 200pt 70pt 200pt, clip, width=0.48\textwidth]{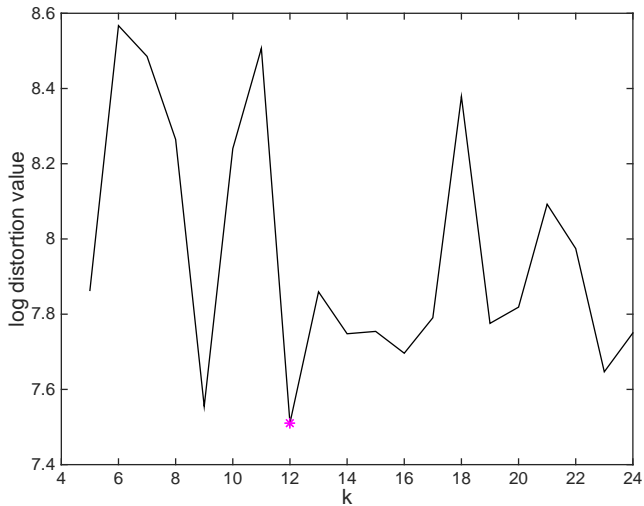}
  \caption{Best trial out of 20: The log distortion values show a local optimum at k = 12 over the range of 5 to 24 clusters (magenta asterisk).}
  \label{fig:kmeans-conv}
\end{figure}

Due to the random initializations, we ran 20 trials for each clustering algorithm in order to obtain the best locally optimal centroids. These optima correspond to 12 and 8 clusters for k-means and DP-means, respectively. All code were implemented in MATLAB and computations executed on a 2.7 GHz Intel Core i7 with 8 GB RAM. Figure~\ref{fig:kmeans-conv} shows an example of the log of distortion values attained over the range of $k$ values for the k-means algorithm ran with 10-fold cross validation. Table~\ref{tab:clus} summarizes the results for the clustering algorithms. The recorded computation times were averaged over the 20 trials, and do not include preprocessing and transforming data into features, etc.

\begin{table}[htbp!]
  \centering
  \caption{Play style clustering summary results}
  \begin{tabular}{lccccc}
    \toprule
    & cross val. & param. range & clusters & cpu time \\ \midrule
    k-means & 10-fold & $k = 5, \ldots, 24$ & 12 & 154.1 s \\
    DP-means & 10-fold & $\lambda = 2.5,2.6,\ldots,4.4$ & 8 & 65.4 s \\
    \bottomrule
  \end{tabular}
  \label{tab:clus}
\end{table}

\subsection{Cluster interpretation}

Surprisingly, our consultations with expert, highly-ranked (top 0.2\% worldwide) LoL players corroborated the correctness of the behavior clusters learned by our algorithms. By checking the centroid values corresponding to each feature and using information about the frequency of in-game characters used for each cluster, these expert players were able map each cluster to a specific gameplay type that they had experienced in-game. This suggests that our clustering were intuitively correct. The mappings determined for the 12-clusters k-means result are as follows. 

\begin{itemize}

  \item \textbf{Ranged physical attacker} Clusters 1, 7, and 9
  \begin{itemize}
    \item Players who maintain distance from fights while dealing high damage with long-range attacks
    \item Players in each cluster differ in risk attitudes, such as whether they attack deeper in enemy territory
  \end{itemize}

  \item \textbf{Ambusher} Clusters 3, 8, 11, and 12
  \begin{itemize}
    \item Players who move stealthily around the battlefield and engage in quick, close-ranged combat
    \item Some players prefer a team oriented style, whereas others prefer a more ``lone wolf'' approach
    \item Includes ``hybrid'' roles with other behavior clusters
  \end{itemize}

  \item \textbf{Team support} Cluster 5
  \begin{itemize}
    \item Players who typically assist ranged physical attackers (healing, cooperative attacks, etc.)
  \end{itemize}

  \item \textbf{Magic attacker} Clusters 6 and 10
  \begin{itemize}
    \item Players who rely on magic-based attacks; as opposed to physical damage in the above clusters
    \item Differ in preference for close- or ranged-combat
  \end{itemize}

  \item \textbf{Miscellaneous} Clusters 2 and 4
  \begin{itemize}
    \item No clear style preference
    \item Differs in skill: either a novice player or prefers an all-around gameplay style
  \end{itemize}

\end{itemize}

Interestingly, we notice from expert analysis that there appears to be a hierarchy of clusters. For instance, clusters 6 and 10 fall under the broader ``magic attacker'' category. This suggests that we might consider other clustering models than k-means or DP-means, as these methods assign each observation to only one cluster. We address this further in Section~V.

\subsection{Cluster visualization with PCA}

\begin{figure}[htbp!]
  \centering
  \includegraphics[trim=70pt 210pt 70pt 250pt, clip, width=0.48\textwidth]{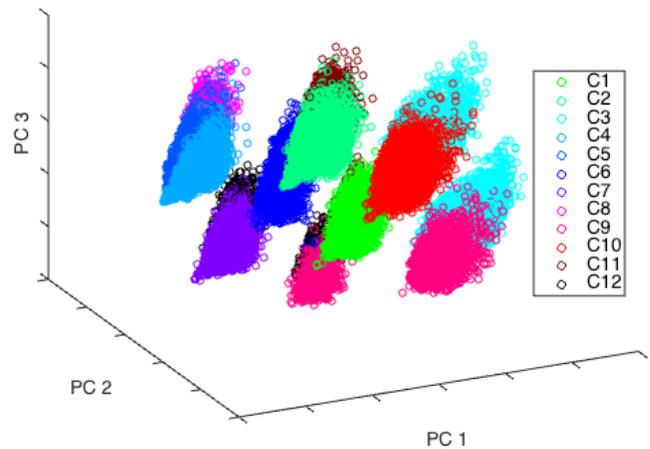}
  \caption{Visualizing our data with 3 principal components reveals at least 8 distinct clusters. The 12-clusters k-means results are overlaid in color.}
  \label{fig:viz}
\end{figure}

Fig.~\ref{fig:viz} shows the result of applying principal component analysis (PCA) to reduce our feature dimension and visualize it in three dimensions. Notice that the data is clearly clustered into 8 distinct groups, suggesting that in higher dimensions there are probably more clusters. Overlaying our 12-groups clustering from the k-means technique in color, we observe that they are consistent with the PCA results: Almost all points in any k-means cluster are in the same PCA cluster. 

\section{Game Outcome Prediction}

We illustrate the accuracy of game outcome predictors that use team composition features based on our gameplay style clusters learned in the previous section. We present our feature selection, the classification models used to learn our predictors, and the accuracies for our predictors. 

\subsection{Feature selection}

For our classification algorithms, the features are the two teams' player compositions. Associated with each team is a vector of counts of players that fall into a certain play style category, which were, say, derived from one of the clustering algorithms. The feature vector is the concatenation of the count vectors of teams 1 and 2. The labels for each sample are the win/loss indicator for the game, with 1 corresponding to a victory and 0 a loss by team 1 to team 2. For instance, there are 8 clusters, teams 1 and 2 have the count vectors $x_{1}\in\reals^{8}$ and $x_{2}\in\reals^{8}$, and team 1 beats team 2. The feature vector-label pair would then be 
\[
  \left(x, y\right) = 
  \left(
    \left[
      \begin{array}{c}
        x_{1} \\ 
        x_{2}
      \end{array}
    \right],
    1
  \right),
\]
where $x$ is the feature vector and $y$ is the sample binary label.

\subsection{Classification models}

To obtain the best win/loss outcome predictor, we consider different classification models. To determine the accuracy of our predictors learned using the various models, we held out 10\% of our total sample set (over 130,000 in total) for training, and used the held-out samples for testing. 

\subsubsection{Logistic regression}

For this model, we use the Bernoulli family of distributions to model the conditional distribution of winning or losing given the team composition features. That is, adhering to our notation introduced above, $y \mid x; \theta \sim \text{Bernoulli}\left(\phi\right)$, where $\theta$ is our model parameter and $\phi = h_{\theta}\left(x\right) = 1/\left(1 + \exp(-\theta^{T}x)\right)$ is our hypothesis, which is derived from formulating the Bernoulli distribution as an exponential family distribution. To learn our model, we find a parameter $\theta$ that maximizes the log-likelihood function
\begin{align}
  \ell\left(\theta\right) & = 
  \log\prod_{i=1}^{m}p\left( y^{(i)} \mid x^{(i)}; \theta \right) \\ & =
  \sum_{i=1}^{m}y^{(i)}\log h_{\theta}\left(x^{(i)}\right) + \left(1 - y^{(i)}\right)\log\left(1 - h_{\theta}\left(x^{(i)}\right)\right),
\end{align}
where $m$ is the sample set size. We used stochastic gradient ascent to efficiently find the optimizer $\theta^{\star}$. 

\subsubsection{Gaussian discriminant analysis}

In this model, we assume that the input features $x$ are continuous-valued random variables and model $p\left(x \mid y\right)$ using a multivariate normal distribution. In other words, we use a generative learning model. In our case,
\begin{align}
  y & \sim \text{Bernoulli}\left(\phi\right) \\
  x\mid y = 0 & \sim \mathcal{N}\left(\mu_{0},\Sigma\right) \\
  x\mid y = 1 & \sim \mathcal{N}\left(\mu_{1},\Sigma\right),
\end{align}
where $\mu_{0}$, $\mu_{1}$, and $\Sigma$ are the means and covariance of the Gaussian distributions. Here, we maximize the log-likelihood of the $m$-samples data
\begin{align}
  \ell\left(\phi,\mu_{0},\mu_{1},\Sigma\right) = \log\prod_{i=1}^{m}p\left( x^{(i)}, y^{(i)}; \phi,\mu_{0},\mu_{1},\Sigma \right).
\end{align}

The result of maximizing $\ell$ with respect to the model parameters is a set of exact analytic equations \cite{Ng:14}, which we compute directly. The derivation of these equations are simple, and we omit them for brevity. 

\subsubsection{Support vector machine}

Assuming that our data are separable with a large ``gap,'' a support vector machine model posits that the size of the geometric margin between some observation point and the decision boundary is proportional to ``confidence level'' that the observation is classified correctly. The result of this model is an optimization problem that seeks the maximum margin separating hyperplane for our samples. 

For our problem, we use $\ell_{1}$ regularization since we are uncertain about whether our data is linearly separable (e.g., outliers, erroneous data). The resulting problem is solved using the sequential minimal optimization algorithm \cite{SBS:98}.

\subsection{Evaluation criteria}

To evaluate the usefulness of game outcome predictor models with features based on our learned behavior clusters, we compare them against a baseline predictor with features based on the game developers' official gameplay classes. As introduced in Section~III, the game developers have grouped the in-game characters into six broad categories, such as assassin or support, which supposedly reflects the character's gameplay style. We learn a logistic regression model with features constructed using these categories and use the 10\% hold-out method for cross validation. 

\subsection{Results and discussion}

To ensure fairness of results, we ran 20 trials for each model to determine the predictor accuracies, which are based on different randomized train and test sets. As with our clustering algorithms, all code were implemented in MATLAB, the computations were executed on a 2.7 GHz Intel Core i7 with 8 GB RAM, and the computation times were averaged over the 20 random trials. Again, these times do not include preprocessing and transforming data into features, etc.

As we observe in Table~\ref{tab:pred}, the best predictor learned using our behavior clusters-based features uses an SVM model with features derived from our k-means clustering. This predictor did significantly better (16\% better) than the baseline algorithm on the test sets, which had 55.1\% and 54.4\% accuracies on the training and testing sets. The other predictors were also competitive---all were only less accurate by a tiny percentage. 

\begin{table}[htbp!]
  \centering
  \caption{Outcome prediction summary results}
  \begin{tabular}{lcccccc}
    \toprule
    & \multicolumn{3}{c}{k-means} & \multicolumn{3}{c}{DP-means} \\
    & train acc. & test acc. & cpu time & train acc. & test acc. & cpu time \\ \midrule
    LR & 72.3\% & 68.8\% & 7.4 s & 69.7\% & 67.1\% & 7.1 s \\
    GDA & 74.8\% & 70.1\% & 7.7 s & 70.9\% & 68.4\% & 7.1 s \\
    SVM & 74.8\% & 70.4\% & 91.2 s & 71.7\% & 69.2\% & 41.6 s \\
    \bottomrule
  \end{tabular}
  \label{tab:pred}
\end{table}

Other than illustrating the relatively high accuracy of our team composition-based outcome prediction approach, this result also implies that our behavior clusters learned had more descriptive power than the official game developers' version. This indirectly concurs with what we have shown from our clustering models: The official gameplay style categories that were used for the baseline algorithm do not necessarily correspond to the behaviors of actual players in games. 

Overall, our results validate our framework of first clustering players by their gameplay style and then using team composition features based on these learned styles to predict team performance. And since our target game is a representative title for games of the same type (i.e., team-based role-playing games), we expect this framework to also be effective and generalizable to other multiplayer games. 

\section{Conclusion and Extensions}

In this brief, we have presented an algorithmic framework for outcome prediction: By learning in-game player behavior categories through clustering and using them in features for game outcome predictors based on classification models, we are able to determine wins and losses with over 70\% accuracy for our target game. This approach could be used to evaluate how team compositions can affect performance in games other than the one we have considered.

Future work will include adding time-dependent player statistics features. Unlike the overall game statistics we used, these timed statistics might give an additional layer of descriptive power, allowing the model to differentiate between clusters based on how players behave early and later in the game. This might lead to a better features for a more accurate team composition-based win/loss predictor. As another extension, we could also consider different clustering models, such as one that captures the ostensibly hierarchical clustering seen in the expert analysis of the k-means results. For instance, the BP-means model described in \cite{TKJ:13} is designed to capture such hierarchical clustering relationships. 

\section*{Acknowledgments}
We thank Professor Andrew Ng and the course staff for motivating and giving feedback for our work. We are also grateful to the LoL expert players who helped with our cluster analysis.

\bibliography{IEEEabrv,report}

\end{document}

%% file: defs.tex
\newcommand{\reals}{{\mbox{\bf R}}}

\newcommand{\argmin}{\mathop{\rm argmin}}

